\documentclass[prb,aps,twocolumn,footinbib,floatfix,10pt,longbibliography,superscriptaddress]{revtex4-1}

\usepackage{chemformula} 
\usepackage[T1]{fontenc} 
\usepackage{graphicx}
\usepackage{dcolumn}
\usepackage{lipsum}
\usepackage{tabu} 
\usepackage{mathtools}

\usepackage{braket}
\usepackage{color}
\usepackage[11pt]{moresize}
\usepackage{anyfontsize}
\usepackage{bbding}
\usepackage{amsmath}
\usepackage{xcolor}
\usepackage{braket}
\usepackage[normalem]{ulem}

\begin{document} 
\author{Patrick Laferri{\`e}re}

\affiliation{National Research Council Canada, Ottawa, Ontario, Canada, K1A 0R6.}
\affiliation{University of Ottawa, Ottawa, Ontario, Canada, K1N 6N5.}
\author{Edith Yeung}
\author{Isabelle Miron}
\affiliation{National Research Council Canada, Ottawa, Ontario, Canada, K1A 0R6.}
\affiliation{University of Ottawa, Ottawa, Ontario, Canada, K1N 6N5.}
\author{David B. Northeast}
\affiliation{National Research Council Canada, Ottawa, Ontario, Canada, K1A 0R6.}
\author{Sofiane Haffouz}
\affiliation{National Research Council Canada, Ottawa, Ontario, Canada, K1A 0R6.}
\author{Jean Lapointe}
\affiliation{National Research Council Canada, Ottawa, Ontario, Canada, K1A 0R6.}
\author{Marek Korkusinski}
\affiliation{National Research Council Canada, Ottawa, Ontario, Canada, K1A 0R6.}
\author{Philip J. Poole}
\affiliation{National Research Council Canada, Ottawa, Ontario, Canada, K1A 0R6.}
\author{Robin L. Williams}
\affiliation{National Research Council Canada, Ottawa, Ontario, Canada, K1A 0R6.}
\author{Dan Dalacu}
\email{Dan.Dalacu@nrc.ca}
\affiliation{National Research Council Canada, Ottawa, Ontario, Canada, K1A 0R6.}
\affiliation{University of Ottawa, Ottawa, Ontario, Canada, K1N 6N5.}

\title{Unity yield of deterministically positioned quantum dot single photon sources}





\begin{abstract}

We report on a platform for the production of single photon devices with a fabrication yield of 100\%. The sources are based on InAsP quantum dots embedded within position-controlled bottom-up InP nanowires. Using optimized growth conditions, we produce large arrays of structures having highly uniform geometries. Collection efficiencies are as high as 83\% and multiphoton emission probabilities as low as 0.6\% with the distribution away from optimal values associated with the excitation of other charge complexes and re-excitation processes, respectively, inherent to the above-band excitation employed. Importantly, emission peak lineshapes have Lorentzian profiles indicating that linewidths are not limited by inhomogeneous broadening but rather pure dephasing, likely elastic carrier-phonon scattering due to a high phonon occupation. This work establishes nanowire-based devices as a viable route for the scalable fabrication of efficient single photon sources and provides a valuable resource for hybrid on-chip platforms currently being developed. 




\end{abstract}

\maketitle 

It has long been recognized that the radiative decay of excitonic complexes in epitaxial quantum dots can produce non-classical photon states e.g. single photons\cite{Gerard_LT1999,Michler_SCI2000,Santori_NAT2002} and entangled photon pairs\cite{Benson_PRL2000,Stevenson_NAT2006,Jayakumar_NC2014}. By incorporating the quantum dots within appropriate photonic structures,  very bright and very efficient sources can be produced \cite{Claudon_NP2010,Somaschi_NP2016,Unsleber_OE2016,Loredo_Opt2016,Wang_PRL2016,Ding_PRL2016}. The cited sources employed randomly nucleated dots formed by strain-driven processes (i.e. self-assembly) and it has been a long term goal to develop techniques to control where such dots nucleate\cite{Chithrani_APL2004, Baier_APL2004,Schneider_APL2008,Strittmatter_PSS2012,Jons_NL2012,Unsleber_Opt2015,Straus_APL2017}. Site-control techniques facilitate the coupling of the emitters to photonic structures\cite{Gallo_APL2008,Schneider_APL2009,Dalacu_PRB2010} and make device manufacture compatible with standard semiconductor processing techniques. They will also be essential in many quantum technologies that rely on the availability of a large number of such sources e.g. linear optical quantum computing\cite{Knill_NAT2001}, quantum simulation\cite{Guzik_NP2012,Sparrow_NAT2018} and Boson sampling\cite{Loredo_PRL2017,Wang_NP2017,Wang_PRL2019}




An approach employing site-control of self-assembled quantum dots that produces high efficiency devices having a high optical quality with a high yield has not yet been realised. In contrast, quantum dots incorporated within nanowires grown using vapour-liquid-solid epitaxy\cite{Borgstrom_NL2005} offer a straight forward approach to site-control\cite{Bjork_NL2002,Dalacu_APL2011}. High efficiency devices are possible using appropriate nanowire geometries designed for single-mode waveguiding operation\cite{Claudon_NP2010}. The bottom-up approach is well suited to providing high efficiency devices as the incorporated dot is naturally aligned on-axis for optimal coupling to the fundamental waveguide mode. Importantly, the processing required for the site-controlled growth does not degrade the optical quality of the emitted photons. Sources based on position-controlled bottom-up nanowires have demonstrated single photon collection efficiencies of $>40\%$ with near-transform limited linewidths of 880\,MHz\cite{Reimer_PRB2016} and single photon purities greater than 99\%\cite{Dalacu_NL2012}. 





\begin{figure*}
\includegraphics*[width=17cm]{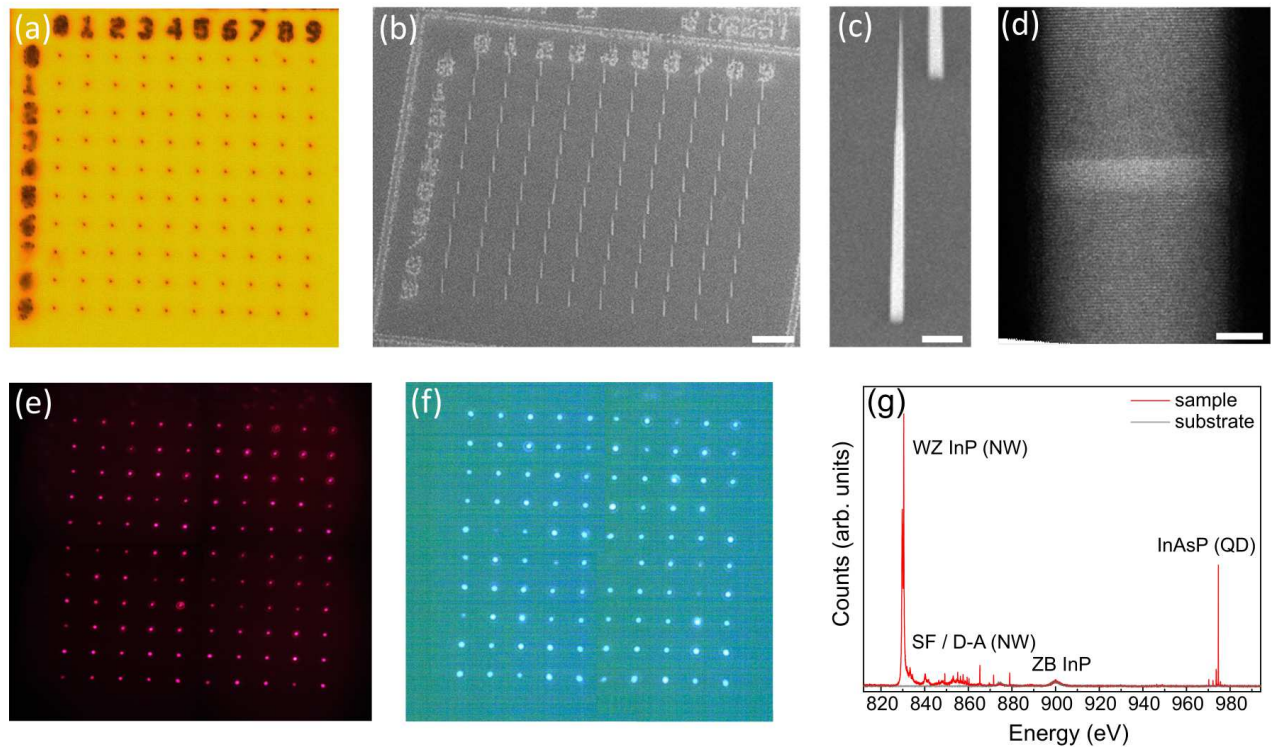}
\caption{(a) Optical image of a 10 $\times$ 10 array of InP photonic nanowires. (b) Scanning electron microscopy (SEM) image viewed at 40$^\circ$ of the same array with (c) showing a higher magnification image of a single nanowire. Scale bars are 10 and 1\,$\mu$m, respectively. (d) Transmission electron microscopy image of the InAsP quantum dot embedded within a nanowire. Scale bar is 5\,nm. Spatially-resolved photoluminescence (PL) from the array in (a) showing emission from (e) the InP photonic nanowire waveguide and (f) the embedded quantum dot. (g) Spectrally-resolved emission from a single device (red curve) showing InP nanowire (NW) and InAsP quantum dot (QD) emission. The NW shows band-to-band emission from wurtzite (WZ) InP as well as emission from stacking faults (SFs) and donor-acceptor (D-A) levels. Emission from the substrate (grey curve) shows two broad features at 875\,nm and 900\,nm associated with band-to-band and D-A recombination, respectively, in zincblende (ZB) InP.}\label{array_images}
\end{figure*}

In the present work, we extend the bottom-up site-controlled approach to realize unity yield of bright and efficient single photon sources. We analyse a large 10 $\times$ 10 array of nanowires and find each and every device displays bright, single emission lines associated with recombination of excitonic complexes from the single quantum dot embedded within. From a detailed characterization of a representative subset of the nanowires in the array, we observe devices that operate with efficiencies that approach theoretical limits given by the calculated dot-mode coupling $\beta \sim 95\%$. A statistical analysis of the performance of the devices afforded by the high yield provides insight into possible mechanisms responsible for preventing universal optimal performance. 



Key properties relevant in quantum technology applications were also measured, in particular single photon and spectral purity. Using above-band excitation we achieve close to optimal performance which, for self-assembled quantum dots, is typically reserved for devices employing charge tunable cavity structures operated under resonant excitation \cite{Kuhlmann_NC2015,Wang_PRL2016,Somaschi_NP2016,Zhai_NC2020}. This surprising result is attributed to having one and only one emitter per device and to the absence of the 2D wetting layer (WL) that is present in self-assembled dots. The former eliminates the detrimental effects of nearby dots (i.e. spectral pollution) which is a persistent issue when employing randomly nucleated dots. The latter eliminates the detrimental effects associated with hybridization of dot and WL states\cite{Lobl_CP2018} (i.e. enhanced exciton-phonon scattering\cite{Urbaszek_PRB2004}).




\vspace{0.5cm}
\textbf{Results}

\textbf{Yield}

The quantum dots studied here are InAsP segments incorporated within site-controlled, bottom-up wurtzite InP nanowires grown using selective-area vapour-liquid-solid epitaxy using gold catalysts\cite{Dalacu_APL2011}. To efficiently collect the emission, the quantum dots are incorporated in photonic nanowire waveguides, see Methods. Optical and scanning electron microscopy (SEM) images of a 10 $\times$ 10 array of nanowires pitched at $7.5\,\mu$m are shown in Fig.~\ref{array_images}(a) and (b), respectively. Each gold catalyst nucleates a well-formed photonic nanowire (Fig.~\ref{array_images}(c)) with a uniform geometry consisting of a 250\,nm diameter base tapered to a 20\,nm tip (size of catalyst) over a length of 10\,$\mu$m. Within each nanowire is a single InAsP quantum dot having a well-defined geometry, shown in the transmission electron image of Fig.~\ref{array_images}(d). 

 
Photoluminescence (PL) measurements were performed with the sample held at 4\,K in a closed-cycle He cryostat. Images of spatially-resolved emission from the array are shown in Figs.~\ref{array_images}(e) and (f). The array is illuminated by a white light source and the PL is imaged onto a CMOS camera.  In (e), a $\lambda=800$\,nm long-wave-pass (LWP) filter is used to filter the white light, and the observed image corresponds to emission predominantly from the InP nanowire (see spectrum in Fig.~\ref{array_images}(g), discussed below). All the nanowires in the array show bright emission with approximately the same intensity. In (f), a LWP filter with a bandedge at $\lambda=950$\,nm is used to filter out both the white light and the InP emission. The image thus corresponds to quantum dot emission. The majority of devices are sufficiently bright to be observed on a simple CMOS camera. 

Fig.~\ref{array_images}(g) shows the spectrally-resolved PL from one of the nanowires in the array. In this case, excitation is from a $\lambda=780$\,nm laser focused onto a single nanowire and the PL is dispersed using a grating spectrometer and detected with a CCD (see Methods). The PL is recorded over a wavelength range that includes both the InP nanowire-related emission and InAsP quantum dot emission.  The InP emission includes band-to-band recombination at $\lambda \sim 832$\,nm as well as longer wavelength emission associated with stacking faults\cite{Akopian_NL2010,Bavinck_NL2012} (i.e. zincblende sections in the predominantly wurtzite crystal structure) and donor-acceptor levels\cite{Dalacu_NL2012}. The InAsP dot emission is observed at $\lambda \sim 975$\,nm, spectrally isolated from the InP emission by $\sim 100$\,nm, allowing for the facile spectral selection employed in Figs.~\ref{array_images}(e) and (f).

\begin{figure}
\includegraphics*[width=8.5cm,clip=true]{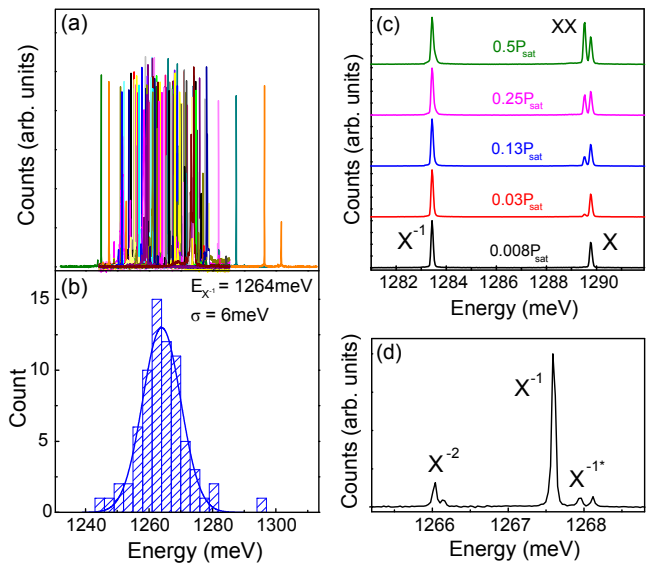}
\caption{(a) Photoluminescence spectra of the 100 nanowires in the array under weak pumping conditions. (b) Histogram of the X$^-$ emission energies extracted from the spectra in (a). (c) Power-dependence of the emission from a typical dot in a nanowire. (d) Example of a dot which shows emission from additional charge complexes.}\label{spectra}
\end{figure}

The PL spectra from all 100 nanowires in the array are plotted in Fig.~\ref{spectra}(a). Excitation was above-band with a $\lambda = 780$\,nm continuous wave (CW) laser using an excitation power, $P$, equal to $\sim 1\%$ of that required to saturate the ground state dot levels, $P_\mathrm{sat}$. The spectrum of each of the 100 dots consists of a few narrow lines and of these 72\% show just two dominant peaks, as in the bottom spectrum of Fig.~\ref{spectra}(c).  We identify these two peaks using excitation power-dependence, cross-correlation spectroscopy\cite{Laferriere_APL2021} and polarization-resolved PL \cite{Reimer_PRB2016}. The high energy peak is associated with emission from the neutral exciton, $X$, whilst the low energy peak with the negative trion (singly charged exciton with both electrons in the s-shell), $X^{-1}$.  The relative probability of emitting a neutral or charged exciton photon, given by the relative peak intensities, is dot-specific and can vary from strongly dominant charged exciton emission to strongly dominant neutral exciton emission, see Ref.~\citenum{Dalacu_PRB2020}. As the excitation power is increased, a third peak typically appears spectrally located between $X^{-1}$ and $X$ which we associate with emission from the biexciton, $XX$. For these 72 nanowires we find most dots have an average charge state corresponding to $X^{-1}$ (i.e. $X^{-1}$ peak is typically brighter than the $X$ peak). Focusing on $X^{-1}$, we show in Fig.~\ref{spectra}(b) the distribution of peak energies extracted from the PL spectra, obtaining a mean energy $E_{X^{-1}}=1264$\,meV with a standard deviation $\sigma =6$\,meV. 

In addition to the 2 dominant lines, approximately 50\% of the 72 devices also show weaker peaks around $X^{-1}$, see Fig.~\ref{spectra}(d). We associate these peaks with recombination from complexes which feature an electron in the p-shell. The peak redshifted with respect $X^{-1}$ is associated with a doubly charged negative trion, $X^{-2}$, whilst the 2 blueshifted peaks correspond to the 2 (doubly generate) bright triplet states of the excited trion\cite{Benny_PRB2012}, $X^{-1*}$. The remaining 28 devices show additional bright peaks suggesting the presence of additional charge states.

\vspace{0.5cm}
\textbf{Count rates}

\begin{figure}
\includegraphics*[width=8.8cm,clip=true]{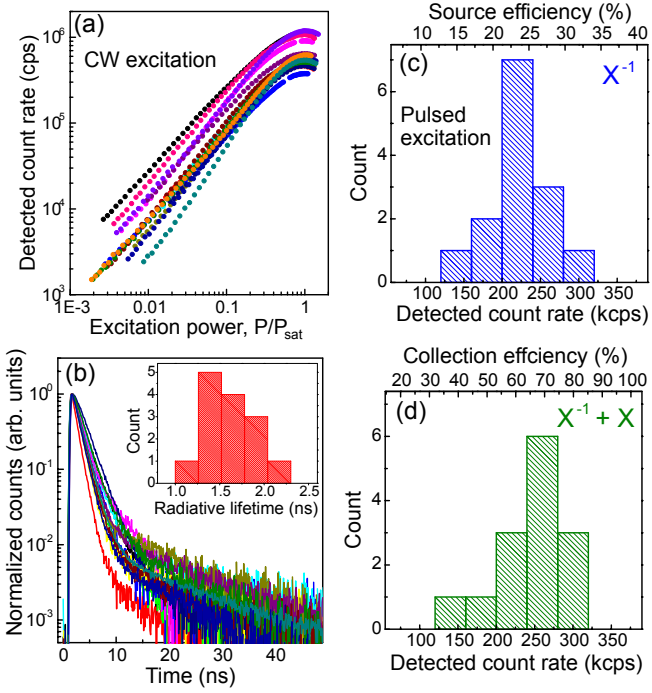}
\caption{(a) Detected count rate of $X^-$ photons as a function of CW excitation power for 14 nanowire devices. (b) Radiative decay rates of the charged exciton complexes in (a). (c) Histogram of detected count rates at P$_{\mathrm{sat}}$ of $X^{-1}$ photons using pulsed excitation at 80\,MHz. Top axis shows source efficiency (i.e. counts at first lens)  (d) As above but counting both $X^{-1}$ and $X$ photons. Top axis shows collection efficiency (i.e. counts from unfiltered photons directed up the nanowire}\label{counts}
\end{figure}

From the 72 devices that showed simple two peak spectra, we selected 14 for further measurements. For these devices, the emission from single excitonic complexes were selected using a tunable 0.1\,nm bandwidth filter and detected with an avalanche photodiode (APD) (see Methods). We chose nanowires for which (1) the dominant charge state corresponded to the charged exciton, as in Fig.~\ref{spectra}(c), and (2) the emission energy of the X$^-$ photon was within the range of our tunable filter ($950 \pm 30$\,nm). The power-dependent count rates obtained from all 14 devices using CW excitation at  $\lambda=780$\,nm are shown in Fig.~\ref{counts}(a). We observe a significant variation in count rates at saturation from device to device, ranging from 0.4 to 1.2\,Mcps. 

For an idealized 2-level system, CW excitation generates the highest count rates achievable, limited by the lifetime of the excited state, here the radiative lifetime of the X$^-$ complex, $\tau_{\mathrm{x}^-}$. To assess the role of lifetime in determining the CW count rate, we measured the PL decay curves of the 14 devices, shown in Fig.~\ref{counts}(b). All the dots display a two-component decay process which we fit with a bi-exponential of the form $\sim a\mathrm{exp}(-t/\tau_{\mathrm{x}^-})+b\mathrm{exp}(-t/\tau_\mathrm{x^{-*}})$ to extract the fast component $\tau_{\mathrm{x}^-}$ (the slow component $\tau_\mathrm{x^{-*}}$ is discussed later in the article). The distribution of extracted radiative lifetimes is shown in the inset of the figure from which we determine an average $\tau_{\mathrm{x}^-} \sim 1.5\pm0.5$\,ns. Comparison of the lifetime-limited emission rates $\Gamma \sim 1/\tau_{\mathrm{x}^-}$ and the CW count rates at $P_\mathrm{sat}$ did not reveal any clear correlation, suggesting that other factors besides the radiative lifetime play a role in determining the ultimate count rate. They are addressed below.



To exclude any count rate variation with radiative lifetime $\tau_{\mathrm{x}^-}$, we measure the count rates at saturation using pulsed excitation with a pulse period of 12.5\,ns, significantly greater than $\tau_{\mathrm{x}^-}$. Detected count rates at saturation from the 14 devices are plotted in Fig.~\ref{counts}(c) and show a narrow distribution around 220\,kcps with $\sigma=39$\,kcps. From these measurements we determine the source efficiency $\eta_s$ (fraction of excitation pulses that produce an X$^-$ photon at the first lens). Using the measured transmission of the optical system ($\eta_t = 7.8\%$) and the detector efficiency ($\eta_d = 15\%$) at $\lambda =980$\,nm, we obtain a mean source efficiency of $23\%$ with a peak value of $30\%$. The distribution of calculated source efficiencies are displayed along the top axis of Fig.~\ref{counts}(c). 

Using these measurements, we also determine the collection efficiency of the devices (fraction of photons emitted from the top of the nanowire that are collected).  We assume only 50\% of the emitted photons are directed towards the tip of the nanowire. Of these, only 80\% reach the detector, assuming 20\% of the photons are emitted from phonon sidebands \cite{Lodahl_RMP2015} and are filtered out by the 0.1\,nm bandpass filter. To obtain an accurate measure of the collection efficiency, all mutually exclusive excitonic complexes that produce photons should be included. Here, we include counts from the two dominant peaks, $X$ and $X^-$, seen in Fig.~\ref{counts}(d). Compared to Fig.~\ref{counts}(c), the distribution is shifted to higher rates and the mean count rate increases to 247\,kcps. Using $\eta_t = 7.8\%$ and $\eta_d = 15\%$ as above, we obtain a mean collection efficiency of 66\% and a peak value of 83\% (see the top axis of Fig.~\ref{counts}(d)).


The peak collection value is lower than expected based on the device design, for which a coupling of the dot emission to the waveguide mode $\beta=95\%$ is predicted \cite{Dalacu_NT2019}. As we do not observe any obvious structural difference between the nanowires from inspection of SEM images and the growth process inherently has the quantum dot positioned on-axis for optimal coupling, we would expect a uniformly high collection efficiency that approaches $\beta=95\%$. One possible explanation for the observed discrepancy is the omission of counts from other excitonic complexes, for example, photons associated with $X^{-2}$ and $X^{-1*}$ recombination seen in Fig.~\ref{spectra}(d).



We also note that collection efficiencies are likely underestimated due to the incoherent excitation employed which can excite dark spin configurations within the dot. For example, the neutral exciton created using above band excitation would be in a dark configuration (i.e. parallel electron and hole spins) 50\% of the time. Similarly, one of the 4 doubly degenerate excited trions states is also dark. These configurations require a spin-flip process to unblock the emission\cite{Johansen_PRB2010,Benny_PRB2014}. Spin-flip rates that are slower than the excitation rate would impact the measured count rate. For the case of the spin-polarized excited trion, a spin-flip and subsequent decay of the p-shell electron produces $X^{-1}$. As such, the spin-flip rate is given by the slow component, $\tau_{\mathrm{x^{-1*}}}$, observed in the decay of $X^{-1}$ (see Fig.~\ref{counts}(b)). Median $\tau_{\mathrm{x^{-1*}}}$ values extracted from fits to the decay curves were $\sim 30$\,ns, i.e. longer than the pulse rate of 12.5\,ns, and varied significantly, from 8\,ns to 120\,ns. These values are comparable to spin-flip rates measured on the neutral exciton in InAs/GaAs self-assembled quantum dots where values of 68 to 215\,ns were reported\cite{Johansen_PRB2010}.

Although devices operating with efficiencies approaching the theoretical maximum can be obtained using above-band excitation, it is clear that the yield of these devices will be limited by the excitation of more than one possible charge and spin configuration. The average charge state can be manipulated to a certain extent using quasi-resonant excitation\cite{Santori_NAT2002} or with additional weak laser excitation at specific above-band energies \cite{Benny_PRB2012}. Alternatively, some degree of control may be possible by manipulating the band-bending at the nanowire surface due to Fermi-level pinning\cite{vanWeert_APL2006} e.g. by tailoring the surface passivation. Improved charge-state control (e.g. via application of an electric field\cite{Baier_PRB2001}) would necessitate modifications to the device design that would allow for the addition of metallic gates.

It appears, however, that consistent (dot-independent) efficiencies may also require coherent pumping of a specific complex using strictly resonant excitation \cite{He_NN2013}. Whether resonant excitation in a charge tunable device will ultimately produce the highest efficiency sources is still unclear and depends on the development of techniques\cite{Ates_PRL2009,Muller_PRL2007,He_NP2019} that address the 50\% loss incurred when using conventional polarization-based rejection of the resonant pump laser \cite{Kuhlmann_RSI2013}. 



\vspace{0.5cm}
\textbf{Single Photon Purity}

\begin{figure}
\includegraphics*[width=9cm,clip=true]{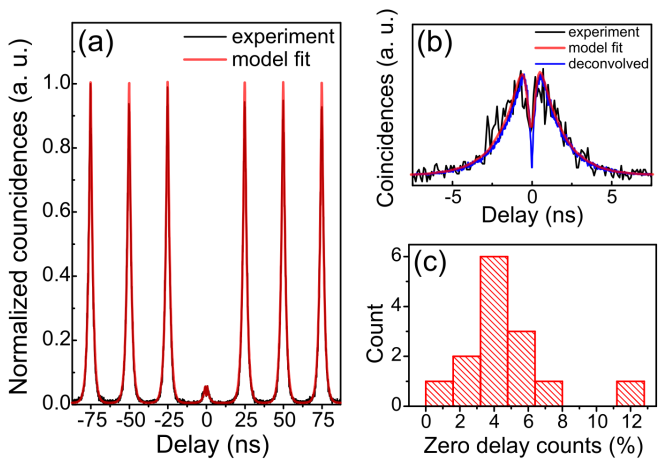}
\caption{(a) Autocorrelation coincidence counts for a device showing re-excitation effects when pumped at saturation. (b) Expanded view of (a) showing the zero delay peak and $\tau=0$ dip. (c) Histogram of the coincidence counts in the zero-delay peak normalized to the counts in the side peaks for the devices pumped at $P=P_{\mathrm{sat}}$.}\label{g2}
\end{figure}

We look next at the single photon purity of the emitted photons from the 14 devices when operated at $P=P_\mathrm{sat}$. The second-order autocorrelation, $g^{(2)}(\tau)$, is measured in a standard Hanbury Brown and Twiss experiment with above-band, pulsed excitation at 40\,MHz, see Methods. A spectrum of the measured coincidences for a device which shows re-excitation (discussed below) is shown in Fig.~\ref{g2}(a). At $\tau=0$ all the devices show $g^{(2)}(\tau=0)\sim 0$ after deconvolution of the spectrum with the 200\,ps jitter in the detector response (see Fig.~\ref{g2}(b)). The absence of coincidence counts at zero delay for all sources, which is observed at $P=P_{\mathrm{sat}}$, is expected from the single emitter nature of the devices i.e. the deterministic growth eliminates the possibility of spectral pollution from nearby quantum dots.

For increasing absolute delay times, coincidence counts are observed to increase rapidly before decaying back to zero with a time constant $\tau_{\mathrm{x-}}$. This is seen clearly in Fig.~\ref{g2}(b) which is an expanded view of Fig.~\ref{g2}(a) around $\tau=0$. Such coincidence spectra are typical when pumping above-band and are a signature of re-excitation processes in which carriers from the excitation pulse repopulate the quantum dot after emission of the first photon\cite{Aichele_NJP2004,Santori_NJP2004}. Subsequent recombination of the carriers produces a second photon that has a corresponding probability of producing a detector click. We simulate the statistics of detection events using a stochastic model which includes a carrier re-capture probability, see Ref.~\citenum{Laferriere_NL2020} for details. The model quantitatively reproduces the measured correlations (red curves in Figs.~\ref{g2}(a,b)) and we extract a time constant associated with re-excitation (i.e. time constant associated with the dip at $\tau=0$) of $\sim 50$\,ps. 

The coincidence counts associated with re-excitation (i.e counts in the zero-delay peak) normalized with respect to the counts in the side peaks are summarized in the histogram shown in Fig.~\ref{g2}(c) for the 14 devices. We obtain a distribution of single photon purities centred around 96\% with extreme values as high as 99.4\% and as low as 88.2\%. As expected from a pump power-dependent re-excitation process, the relative coincidence count rates in the zero delay peak decrease (i.e. purities approach 100\%) as the excitation power is decreased from $P=P_{\mathrm{sat}}$ (not shown, see Ref.~\citenum{Laferriere_NL2020}). 


\vspace{0.5cm}
\textbf{Spectral Purity}

\begin{figure}
\includegraphics*[width=9.cm,clip=true]{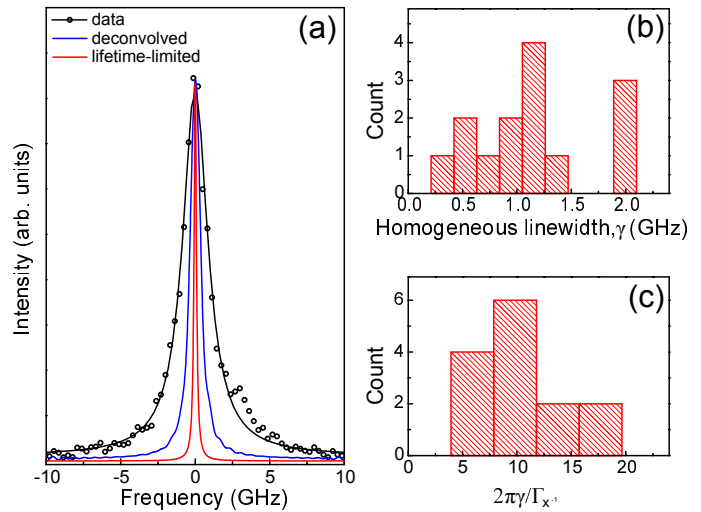}
\caption{(a) Measured linewidth of $X^-$ (symbols) and a Lorentzian fit (black curve). Deconvolved linewidth is $\gamma=0.77$\,GHz (blue curve) which is $8\times$ life-time limit of 0.094\,GHz calculated from measured lifetime (1.7\,ns). (b) Deconvolved linewidths and (c) excess broadening beyond the transform limit extracted from the 14 devices.}\label{linewidths}
\end{figure}

Finally, we measure the spectral purity of the emitted photons from the nanowire devices. The linewidths of the $X^{-1}$ emission peaks were measured using a scanning Fabry-Perot etalon (BW = 1.3\,GHz), see Methods. 13 of the measured lines revealed a Lorentzian profile (black curve in Fig~\ref{linewidths}(a)) having linewidths ranging from 1.68 to 3.28\,GHz.  Only one device displayed an inhomogeneously broadened line profile i.e. the emission line was better described by a Voigt profile with a Gaussian component contributing ~50\% to the linewidth.

 

The absence of inhomogeneous broadening, suggested by the prevalence of Lorentzian lineshapes, demonstrates the high quality of the epitaxial material of the devices. The density of traps in close proximity to the dot, which are typically associated with a fluctuating charge environment responsible for spectral wandering of the emission line \cite{Santori_NAT2002}, appears to be insignificant. We attribute the absence of spectral fluctuation-mediated inhomogeneous broadening to an increase in the growth temperature used during deposition of the InP shell compared to previous experiments \cite{Reimer_PRB2016}. 

The observation of Lorentzian lineshapes allows for direct access to homogeneous linewidths, $\gamma$, from which coherence times $T_2=1/\gamma$ can be calculated. To extract the homogeneous linewidths, the measured spectra were deconvolved from the etalon response (blue curve in Fig~\ref{linewidths}(a)) and are plotted in Fig~\ref{linewidths}(b). The majority of the devices show $\gamma$-values uniformly distributed from 0.4 to 1.5\,GHz with a few lines showing a broader $\gamma$ around  2\,GHz. In the ideal case, where the coherence is governed by population relaxation dynamics and assumed to be purely radiative, one obtains a maximum coherence time $T_2=2T_1$ (here $T_1=\tau_{\mathrm{x}^-}$). Generally, $T_2$ times are limited by pure dephasing mechanisms (e.g. elastic exciton-exciton and exciton-phonon scattering\cite{Scully}) which results in linewidths broadened in comparison to lifetime transform-limited values $1/2 \pi \tau_{\mathrm{x}^-}=\Gamma_{\mathrm{x}^-}/2\pi$ (red curve in Fig~\ref{linewidths}(a)).




To evaluate the broadening in excess of the transform limit, we plot in Fig~\ref{linewidths}(c) the linewidths of the 14 devices in units of $2\pi\gamma/\Gamma_{\mathrm{x}^-}$ using the measured lifetimes $\tau_{\mathrm{x}^-}$ from Fig~\ref{counts}(b). The devices show a wide range of values, from $4\times$ to $20\times$ greater than the transform limit. Although it is not clear why we observe such a large variation, the fact that they are not transform-limited is not surprising. It has been established \cite{Kuhlmann_NC2015} that the attainment of transform-limited single photons from quantum dots not only requires high quality material, but also the use of resonant excitation. With above-band excitation, as is the case here, thermalization of generated carriers to the InP bandedge followed by subsequent capture to the quantum dot levels is expected to generate a high density of phonons. Broadening due to pure dephasing via elastic carrier-phonon interactions is therefore inevitable.


 
\vspace{0.5cm}
\textbf{Discussion}


In summary, we have reported on the performance of nanowire-based single photon sources fabricated using a platform that provides unity device yield. The broadband single emitter device allows for the facile characterization of source performance metrics using above-band excitation. From a sampling of 14 devices, we obtained a peak source efficiency of 30\%, a peak collection efficiency of 83\%, a minimum multiphoton emission probability of 0.6\% measured at saturation and a minimum linewidth of 400\,MHz (maximum coherence time of 2.5\,ns) corresponding to $4\times$ the transform limit. 

From the reliable statistical analysis afforded by the unity yield we quantify the distribution away from optimum efficiencies. The role of emission from more than one possible charge complex is identified as a limiting factor preventing universal peak efficiencies. Achieving uniformly high efficiencies thus entails charge state control using approaches compatible with the nanowire geometry, either on the growth substrate or integrated on-chip\cite{Mnaymneh_AQT2020}, see, for example, Refs. ~\citenum{Zeeshan_PRL2019} and \citenum{Reimer_NL2011}, respectively.

Non-zero multiphoton emission probabilities were associated with a re-excitation process whereas non-transform limited linewidths with phonon-mediated pure dephasing. Both can be traced to the above-band excitation employed. Routes to mitigating these effects are already established and requires coherent pumping of a target charge state.  Additionally, the observation of homogeneous linewidths broadened due to phonon-mediated pure dephasing from the high energy excitation suggests that below-gap pumping should provide significant improvements in the coherence of the emitted photons. In conclusion, we have established nanowire quantum dots as a viable route towards efficiently generating high quality single photons using a process that achieves unity yield. Such devices are a valuable resource for quantum technologies requiring multiple sources. They are particularly relevant in hybrid on-chip platforms utilizing pick and place techniques.


\vspace{0.5cm}
\textbf{Methods}

\vspace{0.5cm}

\textbf{Fabrication and growth} The quantum dots were incorporated in site-controlled, bottom-up nanowires grown by selective-area vapour-liquid-solid epitaxy\cite{Dalacu_APL2011} in the wurtzite InAs/InP material system. The growth was carried out using chemical beam epitaxy on patterned InP substrates with single gold catalysts centered in circular openings in an SiO$_2$ mask (see Ref.~\citenum{Dalacu_NT2009} for details). The first of a two-step growth process produces a nanowire core into which was incorporated a single InAs$_\mathrm{x}$P$_{\mathrm{1-x}}$ dot (Fig.~\ref{array_images}(d)) with a thickness of $\sim 5$\,nm, a diameter of $\sim 20$\,nm and a composition of $\mathrm{x}\sim 25\%$. In the second growth step, the nanowire core was clad with a $\sim 125$\,nm InP shell\cite{Dalacu_NL2012} which defines the photonic nanowire waveguide\cite{Claudon_NP2010} into which the dot emits. For the shell growth, the temperature was increased by $20^\circ$C to $450^\circ$C and the III/V ratio was adjusted to produce the tapered geometry\cite{Dalacu_NM2021} observed in Fig.~\ref{array_images}(c) in order to obtain a an emission profile suitable for efficient collection\cite{Gregersen_OL2008,Bulgarini_NL2014}.

 \vspace{0.5cm}
\textbf{Optical characterization} Photoluminescence (PL) measurements were made in a closed-cycle helium cryostat at 4\,K. The nanowire quantum dots were excited through a $100\times$ objective (numerical aperture\,=\,0.81) using either CW ($\lambda = 780$\,nm) or pulsed ($\lambda = 670$\,nm, pulse width $=100$\,ps) excitation. Emission was collected through the same microscope objective and, for low-resolution (60\,$\mu$eV) spectrally-resolved measurements, was directed to a 0.5\,m grating spectrometer equipped with a liquid nitrogen-cooled CCD array detector. For source efficiency, high-resolution and time-resolved measurements, the emission was fibre-coupled and single photon avalanche photodiodes (APDs, timing jitter $\sim 200$\,ps) were used for detection. A fibre-based tunable filter (bandwidth 0.1\,nm) was used to isolate specific quantum dot emission lines. For second-order correlation measurements, the filtered line was directed to two APDs via a 50:50 splitter in a typical  Hanbury Brown and Twiss (HBT) experiment. The high-resolution measurements were performed by scanning a tunable fibre-based Fabry-Perot etalon (bandwidth 1.3\,GHz) through the selected line. 


\subsection{Data availability}

The data that support the findings of this study are available from the corresponding author upon request.

This work was supported by the Canadian Space Agency through a collaborative project on Development of Quantum Dot Based QKD-relevant Light Sources and by the Natural Sciences and Engineering Research Council of Canada through a Strategic Partnership Grant Quantum-secured Communications for Canada.

\bibliography{whiskers}   
\end{document}